\documentstyle[12pt,epsfig]{article}
\begin{document}

\vskip 1 true cm

\begin{flushleft}
LYCEN 9843
\end{flushleft}

\begin{center}
{\bf CHIRAL SYMMETRY RESTORATION}
\vskip 0.8 true cm
{\bf AND PION INTERACTION IN NUCLEAR MATTER}
 
\vskip 1 true cm
 
Guy CHANFRAY and Dany DAVESNE 
\vskip 0.5 true cm
{\it IPN-Lyon, 43 Bd. du 11 Novembre 1918,
F-69622 Villeurbanne C\'{e}dex, France.}
\end{center}
\vskip 1true cm

\begin{center}
{\bf Abstract}
\end{center}

{\it This paper is devoted to the interplay between p-wave, s-wave 
pion-nucleon/nucleus interaction and in-medium pion-pion interaction 
with special emphasis on the role of the nuclear pionic scalar 
density driving a large amount of chiral symmetry restoration. 
In particular we show that the $\pi NN$ coupling constant and the 
Goldberger-Treiman relation are preserved in the nuclear medium 
within certain conditions.  
We also discuss the related problem of the in-medium pion-pion 
strength function.}

\section{Introduction}

The collective pionic modes originating mainly from the coupling of the 
pion to delta-hole configuration have suscitated considerable interest. 
As proposed in \cite{CHAE1} and confirmed by detailed calculations 
(\cite{GUI,UDA}), the direct experimental evidence of their existence 
even at peripheral nuclear density is provided by charge 
exchange reactions \cite{CON,HEN}. These collective modes, sometimes 
called pisobars, are expected to play a prominent role in highly excited 
hadronuclear matter produced in relativistic heavy ion collisions. Indeed, 
it has been predicted \cite{CHAS,ASA,HERR} that they may significantly affect 
the rho meson mass distribution at large density. As shown in \cite{CHAW1,RAPP},
they are able to explain an important part of the excess of dilepton 
production rate in the $500$ MeV invariant mass region observed 
by the CERES \cite{CERES}
and HELIOS \cite{HELIOS} collaboration at CERN/SPS. In addition 
it has also been predicted
 \cite{CASN} an important 
reshaping of the pion-pion interaction 
in the scalar-isoscalar channel (``sigma'' channel), producing 
a sizeable accumulation of strength near or even below two-pion threshold.
A detailed discussion of this problem, has been given in \cite{AOUSS}. In 
particular, the influence of chiral symmetry constraints, stopping 
the subthreshold invasion of strengh due to the coupling of the pion 
to nucleon-hole states, have been extensively investigated. It has been 
soon realized that this medium effect is of considerable importance for the
still open problem of nuclear saturation since an important part 
of the nucleon-nucleon attraction comes from correlated two-pion exchange 
and several papers have brought extremely interesting results 
\cite{DKW,RDW,RMDB}. However,
this in-medium modification  of the $N N$ interaction is obviously 
very sensitive to the precise low mass shape of the in-medium $\pi\pi$ 
strength function. In that respect Oset {\it et al.} 
have drawn the attention on vertex 
correction on in-medium pion-pion interaction imposed by the standard 
chiral lagrangian \cite{WROS}. 
According to this paper some strength should appear 
below the two-pion threshold at densities below normal nuclear matter density.

One of the common features of the problems mentionned above, is the interplay 
between p-wave pion-nucleon/nucleus interaction, s-wave pion-nucleon/nucleus 
interaction and in-medium pion-pion-interaction all being related to the amount 
 of chiral symmetry restoration in the nuclear medium.
The aim of this paper is to clarify this important question in a 
unified scheme based on the standard chiral lagrangian treating simultaneously
pion propagation, pion-pion interaction and chiral symmetry restoration in the 
nuclear medium. In section 2, we show how the pion scalar density $<\Phi^2>$ 
(directly related to the pion cloud contribution of
the pion-nucleus sigma term governing chiral symmetry restoration) 
may {\it a priori} 
renormalize the $\pi NN$ coupling in the nuclear medium. Starting 
from the in-medium modified pion-pion interaction, we discuss
in section 3 how this scalar density governs the s-wave $\pi-nucleus$ 
interaction through 
the interaction of the external pion 
 with the nuclear pion cloud; we also give an explicit expression 
(which slightly differs from \cite{CHAE2}) of the pion-nucleus sigma commutator 
in terms 
 of the full longitudinal spin-isospin response function. We discuss the 
consequences on the pion propagator in section 4, showing in particular that 
the p-wave $\pi NN$ coupling constant is actually unmodified in the medium to 
one pion loop order. Finally we demonstrate in section 5 that the presence 
of the $3\pi NN$ interaction cannot modify the pion-pion strength function
at finite density at variance with the conclusion of \cite{WROS}.

\section{The chiral lagrangian }
All the results will be derived within a chiral $SU(2)\times SU(2)$ 
non linear lagrangian with anexplicit chiral symmetry breaking 
piece containing a nucleon 
field $N$ and a isovector pion field $\vec \Phi$ 
embedded in the  $2\times 2$ matrix~:
\begin{equation}
U=\xi^2=e^{i \vec\tau.\hat\Phi F(X)},\qquad X^2={\Phi^2\over f_\pi^2},\qquad
F(X)=X+\alpha X^3+\alpha' X^5+...\label{REP}
\end{equation}
where $F(X)$, which is an odd function of $X$, defines the particular non linear 
representation. The most usual choices are the PCAC choice 
($F=\arcsin X$,  $\alpha=1/6$), the Weinberg choice
($\alpha=-1/12$) and the choice $F=X$ commonly used in chiral pertubation expansion. Since
the physical observables should not depend on the representation we will 
not specify it as a consistency check of the results. In the following 
we will treat the soft pion interaction to one pion loop order 
and a practical consistency 
test will be the independence of our results with respect
to the parameter $\alpha$. The lagrangian is~:

\begin{equation}
{\cal L} = {f^2_\pi\over 4}\,tr \partial_\mu U \partial^\mu U^\dagger\,+\,
{1\over 4}  f^2_\pi m^2_\pi \,tr(U+U^\dagger)
\, +\,i\bar N\gamma^\mu\partial_\mu N \,-\,M_N  \bar N  N\,
\,+ \bar N\gamma_\mu {\cal V}^\mu_c N\,
\, + g_A\, \bar N\gamma_\mu\gamma^5 {\cal A}^\mu_c N \nonumber
\end{equation}
with
\begin{equation}
{\cal V}^\mu_c={i\over 2}\left(\xi\partial_\mu\xi^\dagger +
\xi^\dagger\partial_\mu\xi\right)\,\qquad
{\cal A}^\mu_c={i\over 2}\left(\xi\partial_\mu\xi^\dagger -
\xi^\dagger\partial_\mu\xi\right) \label{LAG1}
\end{equation}
To lowest order the pion-pion interacting lagrangian is~:
\begin{equation}
{\cal L}_{4\pi}={1\over f^2_\pi}\,\left[-m^2_\pi 
\left(\alpha-{1\over 24}\right)\,\Phi^2\,\Phi^2\,
+\,\left(\alpha-{1\over 6}\right) \,\Phi^2\, \partial_\mu\vec\Phi.
\partial^\mu\vec\Phi\,+\,\left(2\alpha +{1\over 6}\right)
\vec\Phi.\partial_\mu\vec\Phi\,\vec\Phi.\partial^\mu\vec\Phi\right]
\label{L4PI}
\end{equation}
The $\pi NN$  and the $3\pi NN$ pieces are~: 
\begin{equation}
{\cal L}_{\pi NN} = {g_A\over f_\pi}\,\bar N \gamma^\mu\gamma^5 
{\vec\tau\over 2}.\partial_\mu\vec\Phi N 
\end{equation} 
\begin{equation}
{\cal L}_{3\pi NN} = {g_A\over f_\pi^3}\,\bar N \gamma^\mu\gamma^5 
{\vec\tau\over 2}.\left[\left(\alpha-{1\over 6}\right)\Phi^2
\partial_\mu\vec\Phi  
+\left(2\alpha +{1\over 6}\right)\vec\Phi\,\vec\Phi.\partial_\mu\vec\Phi
\right] N \label {LPINN}
\end{equation}
It follows that, to one pion loop order, the in-medium effective 
$\pi NN$ lagrangian becomes~: 
\begin{equation}
{\cal L}^{eff}_{\pi NN} = {g_A\over f_\pi}\,\left(1+\beta 
{<\Phi^2>\over 6 f_\pi^2}\right)
\,\bar N \gamma^\mu\gamma^5 
{\vec\tau\over 2}.\partial_\mu\vec\Phi N,\qquad 
\beta=1+10\left(\alpha-{1\over 6}\right) \label{LEFF}
\end{equation}
In principle the expectation value of $\Phi^2$ renormalizing a nucleon 
vertex  does not correspond to its averaged value due to the presence of 
short range correlations. This effect has been discussed in a recent paper 
\cite{CDE2} and found to be moderate because 
the scalar pion cloud of a given nucleon 
is mainly outside of the correlation hole. In the following, we disregard 
this effect of short range correlation and calculate 
the pion loop $<\Phi^2>$ with the in-medium pion propagator
including the effect of the p-wave interaction from 
${\cal L}_{\pi NN}$. The retarded pion propagator $D_R$ and the time ordered 
pion propagator $D_R^{++}$ have the form~:  
\begin{eqnarray}
D_R(q) &=& 
\left(q_0^2-\omega^2_q-{\bf q}^2 
\tilde\Pi^0(q_0,{\bf q})\right)^{-1}\nonumber \\
D_R^{++}(q) &=& (1+f(q_0)) D_R(q)-f(q_0) D^*_R(q) \label{PROPR} 
\end{eqnarray}
from which we get~: 
\begin{equation} 
<\Phi^2>=3\int{i\, d^4 k\over (2\pi)^4}\, e^{i k_0\eta_+}\,
 D_R^{++}(k)\,=3\,\int {d^3 k\over (2\pi)^3}\,
\int_0^\infty dE \,\left(-{2E \over \pi}\, Im D_R (E,{\bf k})\right)\,
{1+2 f(E)\over 2E}
\end{equation}
where the second form incorporating the thermal factor 
($f(E)=(exp(\beta E)-1)^{-1}$) can be used at finite temperature if needed. 
In the previous expressions the contribution of the free space propagator 
is understood to be always substracted.
In the following we will work most of the time at zero temperature and make 
no distinction between time ordered and retarded propagators which coincide 
for positive frequencies. All the self-energies and related quantities 
will be expressed in terms of time ordered propagators as usual. 
It is nevertheless interesting to remark that 
most of  the formal results will be valid at finite temperature by replacing 
the energy integral by Matsubara frequencies summation, propagators by 
Matsubara propagators and finally taking the analytical continuation 
($i\omega_n \to \omega+i\eta$) to obtain the corresponding self-energy.
 With this lagrangian,
the pion polarizability is $\tilde \Pi^0=(g_A/2f_\pi)^2 U_N$ where 
$U_N$ is nothing but the standard nucleonic Lindhardt function 
(with an explicit factor $4$ for spin-isospin). 
However to make contact
with the well established phenomenology of nuclear collective pionic modes,
one can incorporate as usual the effect of 
the delta as well as the screening effect from short range correlations ($g'$ 
parameters); details can be found in many papers, see e.g. \cite{CHAS, AOUSS}.
\section{s-wave pion-nucleus interaction and  pion-nucleus sigma commutator}
In free space, the $\pi\pi$ interaction potential in the $I=0$ channel 
can be straightforwardly obtained from ${\cal L}_{4\pi}$~:
\begin{equation}
<q_a q_b|{\cal M}_0|q_c q_d>={1\over f_\pi^2 }\,\left(m^2_\pi-2 s\,+\,\beta
\sum_{i=1}^4 (m^2_\pi-q_i^2)\right),\qquad
\beta =1+10\left(\alpha-{1\over 6}\right)\label{M0}
\end{equation}
where $s=(q_a+q_b)^2=(q_c+q_d)^2$  is the center of mass 
energy of the pion pair. For on-shell pions ($q^2_i-m^2_\pi=0$) 
the $\alpha$ dependent piece proportional to the inverse pion 
propagators disappears as it should be. When two pions are soft (i.e. 
$q_a=q_c=0$), in the PCAC representation 
($\alpha=1/6$), this amplitude satisfies  the soft pion theorem 
\begin{equation}
{\cal M}_0\,(s=u=m^2_\pi, t=0)\,= {m^2_\pi\over f_\pi^2}={\Sigma_{\pi\pi}
\over f_\pi^2}
\end{equation}
where $\Sigma_{\pi\pi}=m^2_\pi$ is the pion-pion sigma commutator 
(with non relativistic normalization for pion states its value is $m_\pi/2$).
This result has to be compared with the Weinberg results for on-shell 
pions at threshold  ${\cal M}_0\,(s=4 m^2_\pi,u= t=0)= -7 m^2_\pi/f_\pi^2$. The 
very important practical consequence of these 
chiral constraint is that the potential, although 
attractive in the threshold region, has to become repulsive somewhere below,
thus preventing the invasion of strength and s-wave pion pair condensation 
in the nuclear medium \cite{AOUSS}.

In the medium this pion-pion potential receives vertex corrections
 depicted on fig.1, which can be calculated from ${\cal L}_{\pi NN}$ and
 ${\cal L}_{3\pi NN}$. For a zero momentum     pion pair (${\bf P}=0$),               
the effective $I=0$ in-medium pion-pion potential takes the very simple form~:
\begin{equation}
<q_a q_b|{\cal M}^{eff}_0|q_c q_d>\,=\,
{1\over f_\pi^2 }\,\left(m^2_\pi-2 s\,+\,\beta
\sum_{i=1}^4 \left( m^2_\pi-q_i^2+ {\bf q}^2_i\,\tilde\Pi^0(\omega_i,{\bf q}_i)
\label{MEFF}
\right)\right) 
\end{equation}
Hence, the effect of the vertex corrections depending on the p-wave pion 
polarizabilities at each pion leg is to make the quasi-potential 
representation independent ({\it i.e.} $\alpha$ independent) for on-shell
quasi-pions satisfying $q^2_i-m^2_\pi-{\bf q}^2_i\,
\tilde\Pi^0(\omega_i,{\bf q}_i)=0$.

The s-wave isoscalar pion 
scattering amplitude in the medium receives a contribution 
from the   interaction of the external pions with the  pion cloud. 
Using crossing symmetry,
this amplitude can be expressed in terms of the previous effective 
interaction~:
\begin{eqnarray}
<q_1|T|q_2> &=& T(q^2_1, q^2_2, t)=
{V\over 2 }\,\int {i d^4 k\over (2\pi)^4}\, D_R(k)\,
<q_1,-q_2|{\cal M}^{eff}_0|k,-k>\nonumber\\
&=&{V\over 2 f^2_\pi}\,\int {i d^4 k\over (2\pi)^4}\, D_R(k)\,
\left(m^2_\pi-2t -\beta\,(2 D^{-1}_R(k)+q^2_1-m^2_\pi + q^2_2 -m^2_\pi)\right)
\end{eqnarray}
where $t$ is the Mandelstam variable $t=(q_1-q_2)^2$ and $V$ the volume 
of the hadronic matter. The contribution of the inverse pion propagators 
disappears  once the vacuum contribution is substracted. Let us 
consider a piece of nuclear matter with $A$ nucleons and density $\rho=A/V$; 
the pion cloud contribution to the pion-nucleus sigma term per nucleon is 
\cite{CHAE2}~:
\begin{equation}
\Sigma_A^{(\pi)}={ m^2_\pi\over 2\rho} <\Phi^2> +{\cal O}(<\Phi^4>)=\, 
{ 3 m_\pi^2\over 2 \rho} \, \int {i d^4 k\over (2\pi)^4}\, D_R(k) \label{SIGMA1}
\end{equation}
Hence, in  the PCAC scheme ($\beta=1$), 
this  pion amplitude (per nucleon) takes the form~:
\begin{equation}
{1\over A} T(q^2_1, q^2_2, t)={\Sigma_A^{(\pi)}\over f_\pi^2 m^2_\pi}
\left(m_\pi^2-t+{1\over 3}(t-q^2_1-q^2_2)\right)
\end{equation}
and satisfies the well known PCAC constraints for 
$\nu_B=(t-q_1^2-q_2^2)/4 M_N=0$; namely~:
soft point~: $T(0,0,0)=\Sigma_A / f^2_\pi$; Cheng-Dashen point~:
$T(m^2_\pi,m^2_\pi, 2 m^2_\pi)=-\Sigma_A / f^2_\pi$; 
Adler consistency relation~: 
$T(m^2_\pi, 0, m^2_\pi)=T( 0, m^2_\pi, m^2_\pi)=0$.
 One may notice that the amplitude at threshold is one third of the 
soft pion point. In particular, in the nucleon case, this gives an isospin 
symetric scattering length $a_0=-\Sigma^{\pi}_N/12\pi f^2_\pi$; taking 
for the pion cloud contribution to the pion-nucleon sigma 
commutator a value of about $30$ MeV (see \cite{JTC,BMG} and discussion below)
one obtains $a_0\approx -0.012 m_\pi^{-1}$ which is rather close 
to the experimental value. However this agreement is probably accidental 
since many other effects may contibute to the scattering length. 
In first rank the Born term alone also gives approximatively the 
experimental value;  the valence quark contribution to the sigma term,
the delta contribution should be also incorporated; similarly in the nuclear
case, effect of rescattering of the isovector amplitude 
in presence of Pauli correlation are known to be important. Since 
s-wave scattering length is not the purpose of this paper we do not 
elaborate on it, although our approach might be implemented 
in this direction along the lines of \cite{DCE}.\par
\noindent
The important quantity to be evaluated is $<\Phi^2>/f^2_\pi=2\rho 
\Sigma^{(\pi)}_A/f^2_\pi m^2_\pi$. This pion cloud contribution to the
nuclear sigma term has already been obtained in \cite{CHAE2} with 
a different method. Here we derive its explicit form using the analytical
structure of the pion propagator in eq.\ref{SIGMA1}. 
Let us introduce the full longitudinal
spin-isospin response function related to the imaginary part 
of the full pion polarization propagator $\Pi_L$ according to~:
 
\begin{eqnarray}
R_L(\omega, {\bf k}) &=&-{V\over \pi} Im\,\Pi_L (\omega, {\bf k})=
-{V\over \pi} Im \left( {\bf k}^2\tilde\Pi_0(\omega, {\bf k})\,
{\omega^2-\omega^2_k\over \omega^2-\omega^2_k -
{\bf k}^2\tilde\Pi_0(\omega, {\bf k})}\right)\nonumber \\
&=& 3 \left({g_{\pi NN}\over 2 M_N}\right)^2\, v^2({\bf k})\,
\sum_n\left|<n| \sum_{i=1}^A\,\vec\sigma(i).{\bf k} \,\tau_{\alpha}(i)
\,e^{i {\bf k}.{\bf x}(i)}|0>\right|^2\,\delta(E_n-\omega)
\end{eqnarray}
where $g_{\pi NN}=M_N g_A/f_\pi$ and $v({\bf k})$ are the $\pi NN$ 
coupling constant and form factor.
Once the vacuum contribution is substracted $<\Phi^2>$ can be calculated 
using a dispersion relation for $\Pi_L$~:
\begin{eqnarray}
<\Phi^2> &=& 3\,\int\,{i\, d^4k\over (2\pi)^4}\,\bigg(D_R(k)-D_0(k)\bigg)\,=\,
3\,\int\,{i\, d^4k\over (2\pi)^4}\,\left({1\over k_0^2-\omega_k^2 
+i\eta}\right)^2\,\Pi_L(k)\nonumber\\
&=& 3\,\int\,{d^3 k\over(2\pi)^3}\,\int_{-\infty}^{+\infty} 
{i dk_0\over 2\pi}\,\left({1\over k_0^2-\omega_k^2 
+i\eta}\right)^2\,\int_0^{+\infty} d\omega \,
\left({-2\omega\over \pi}\right)\,{Im \Pi_L(\omega, {\bf k})\over
k_0^2-\omega^2 +i\eta}
\end{eqnarray}
Exchanging the ordering of the $k_0$ and $\omega$ integrations and performing 
a standard contour $k_0$ integration, we finally obtain for  
$\Sigma_A^{(\pi)}$~:
\begin{equation}
\Sigma^{(\pi)}_A={3 m^2_\pi\over 2 A}\,\int{d^3 k\over (2 \pi)^3}\,
\int_0^\infty d\omega\,\left({1\over 2\omega^2_k (\omega+\omega_k)^2}\,+
\,{1\over 2\omega^3_k (\omega+\omega_k)}\right)\,R_L(\omega, {\bf k}) 
\label{SIGMA2}
\end{equation}
Taking the zero density limit, we recover the pion cloud contribution 
to the pion-nucleon sigma term whose explicit form (ignoring the delta width) 
 is~:
\begin{eqnarray}
\Sigma^{(\pi)}_N &=&{3 m^2_\pi\over 2}\,\int{d^3 k\over (2 \pi)^3}\,
\left({g_{\pi NN}\over 2 M_N}\right)^2\, v^2({\bf k})\,
\left[\left({1\over 2\omega^2_k (\epsilon_k+\omega_k)^2}\,+
\,{1\over 2\omega^3_k (\epsilon_k+\omega_k)}\right)\right.\nonumber\\
& +&{16\over 9}\,\left({g_{\pi NN}\over g_{\pi N\Delta}}\right)^2
\,\left.\left({1\over 2\omega^2_k (\omega_{\Delta k}+\omega_k)^2}\,+
\,{1\over 2\omega^3_k (\omega_{\Delta k}+\omega_k)}\right)\right]\label{NUC}
\end{eqnarray}
with $\epsilon_k=k^2/2 M_N$ and $\omega_{\Delta k}=M_\Delta -M_N + 
k^2/2 M_\Delta$. $\Sigma^{(\pi)}_N$ has already been evaluated within chiral 
quark models like CBM \cite{JTC}  providing a form factor from the nucleon size 
and specific value for $g_{\pi NN}/g_{\pi N\Delta}$. It has been found 
of the order of $30$ MeV with about $10$ MeV from the delta sector, 
the rest ($15$ MeV) coming from the scalar quark density inside the nucleon.
Now, the interesting question is how much the $\pi$-nucleus 
sigma term per nucleon deviates from the $\pi$-nucleon one or, 
said differently , what is the pion exchange contribution~? This problem has 
already been adressed in \cite{CHAE2}; let us briefly summarize the 
conclusion of this work. The nucleon-hole sector contribution 
can be reasonnably calculated within a static approximation, i.e. putting
$\omega=0$ in the prefactor of eq.(\ref{SIGMA2}); the energy integration yields 
a two-body spin-isospin ground state matrix element once the free 
nucleonic part has been substracted. According to a calculation with
correlated wave functions, it turns out that the Pauli blocking 
effect is almost exactly compensated by the effect of tensor 
correlations. Hence, the nucleonic sector itself does not contribute 
to the modification of the sigma term. For the delta sector, the situation 
is less clear. The occurence of collective pion-delta states (the so-called
pionic branch) shifts part of the strength at lower energy and 
one can expect an increase of the sigma commutator closely related 
to the pion excess in nuclei. This expected feature is 
consistent with a dispersive analysis a la Fubini-Furlan which gives 
an increase of the sigma term of about $5$ MeV. However the 
full calculation of eq.(\ref{SIGMA2}) with a realistic  longitudinal
spin-isospin response function, including $p-h$, $\Delta-h$ and $2p-2h$ 
sectors remains to be done \cite{MAR}. 
\section{Pion propagator and in-medium pion-nucleon coupling constant}
In addition to the usual p-wave piece, the pion self-energy
receives a contribution of the pion loop which contains the effect
of the modification of the $\pi NN$ vertex present 
in the pion-pion quasi-potential. It reads~:
\begin{equation} 
S(\omega,{\bf q})={\bf q}^2 \tilde\Pi_0(\omega, {\bf q})\,+\,
{<\Phi^2>\over 6 f^2_\pi}\,
\left[m^2_\pi-2\beta\left(\omega^2-\omega^2_q-{\bf q}^2 
\tilde\Pi_0(\omega, {\bf q})
\right)\right]
\end{equation}
The full inverse pion propagator with pion loop effect takes the form~:
\begin{eqnarray} 
\tilde D^{-1}(\omega,{\bf q})&=&\omega^2-\omega^2_q\,-\,S(\omega,{\bf q})
\nonumber\\
&=&\left(1+\beta {<\Phi^2>\over 3 f^2_\pi}\right)
\,\left(\omega^2-\omega^2_q-{\bf q}^2 \tilde\Pi_0(\omega, {\bf q})
\right)\,+{<\Phi^2>\over 6 f^2_\pi}\,m^2_\pi
\end{eqnarray}
As expected from the effective $\pi NN$ effective lagrangian 
of eq. (\ref{LEFF}), the p-wave self enegy is modified, in this
one-pion loop approximation 
by a factor $1+\beta<\Phi^2>/3 f^2_\pi\approx (1+\beta<\Phi^2>/6 f^2_\pi)^2$.
However, as we will see just below, 
a wave-function renormalization will just compensate this effect.
To clearly see what is going on, we drop the pure s-wave part of the 
self-enegy and concentrate on the p-wave part. In this case the 
full pion propagator becomes~:
\begin{equation}
\tilde D (\omega,{\bf q})=\gamma\,  D_R(\omega,{\bf q})=
{\left(1+\beta {<\Phi^2>\over 3 f^2_\pi}\right)^{-1}\over
\omega^2-\omega^2_q-{\bf q}^2 \tilde\Pi_0(\omega, {\bf q})} 
\end{equation} 
The effective $\pi NN$ coupling constant is modified by the vertex correction 
of (\ref{LEFF}) and the wave-function renormalization factor $\sqrt\gamma$ 
which compensate each other to one pion loop order, namely~:
\begin{equation}
{g_{\pi NN}^*(\rho)\over g_{\pi NN}}= \sqrt\gamma\,\,
\left(1+\beta {<\Phi^2>\over 6 f^2_\pi}\right)=\, 1\,+\, {\cal O}(<\Phi^4>)
\end{equation} 
Hence, to one-pion loop, the pion-nucleon coupling constant is 
unmodified in the nuclear medium. Similarly, if we consider pion exchange 
between two nucleons the two vertex renormalizations exactly compensate 
the $\gamma$ factor of the propagator. Consequently, the effective pion 
propagator, whose imaginary part is directly related to the full
pionic response function, is just the usual $D_R$ calculated 
with an irreductible pion self-energy $\Pi_0$ which is also unmodified 
by pion loop effects. This result justifies a posteriori all
the phenomenological studies of pion-nucleus interaction \cite{BIBLE}
and charge exchange reaction \cite{CHAE1, GUI, UDA}. 
The occurence of this $\gamma$ factor has been previously established
in \cite{CEW} at finite temperature for a hot pion gas 
for which $<\Phi^2>=T^2/4$ in the chiral limit. Our result 
is also valid in this case and one can conclude that the 
$\pi NN$ coupling constant is also unmodified to order $T^2$
as  previously stated in \cite{EK}.

In  recent papers \cite{CDE2, CEW}, we established, 
within the same chiral lagrangian, that the axial coupling constant $g_A$
and the pion decay constant $f_\pi$ are in-medium modified according to~:
\begin{equation}
{f_\pi^*\over f_\pi}={g_A^*\over g_A}=1-{<\Phi^2>\over 3 f^2_\pi} 
\end{equation}
At finite temperature, in a hot 
pion gas, this quenching effect acting on the axial current 
is accompanied by the mixing of the vector and axial correlators 
as demonstrated in \cite{EI}. Since $g_{\pi NN}$ \cite{EK} and the nucleon mass
\cite{LEU} are unmodified, 
we come to the conclusion that the Goldberger-Treiman
relation $M_N \,g_A=g_{\pi NN} \,f_\pi$ is preserved at finite temperature,
as already pointed out in \cite{EK}. 
At finite density the quenching of $g_A$ and $f_\pi$ is also linked to 
a mixing effect intimately related to chiral symmetry restoration 
\cite{CDE2}. 
The novel feature  is that 
the unmodified Goldberger-Treiman relation also holds 
at finite density in the nuclear medium since 
there is no modification of the nucleon mass apart 
higher order effects associated to 
the p-wave dressing of the nucleonic pion cloud.
However we stress that this 
result is strictly valid with neglect of short-range correlations 
affecting the nucleon observables $g_A$ and $g_{\pi NN}$. In \cite{CDE2}
we have shown that the  renormalization of $g_A$ (which is a pure 
vertex renormalization)  is accounted 
for by simply replacing the averaged pion scalar density $<\Phi^2>$
by an effective one incorporating the effect of short-range correlation; 
the net result is a $10\%$ quenching at normal nuclear density. The problem
of the effect of short-range correlation on $g_{\pi NN}$ remains to be 
clarified.

\section{Pion-pion scattering in the nuclear medium}

We are now in position to study pion-pion scattering both in free space and in 
the nuclear medium. In particular,
we want to investigate to which extent the medium modification of 
the $\pi \pi$ interaction, involving the pion polarizability $\Pi_0$
 (eq.(\ref{MEFF})),
influences the two-pion strength function.

Let us first consider the free case; the unitarized 
scattering amplitude  in the 
scalar-isoscalar channel can be obtained 
as the solution of a Lippmann-Schwinger equation~:
\begin{equation}
<q_a q_b|{\cal T}_0|q_c q_d>=<q_a q_b|{\cal M}_0|q_c q_d>\,+\,
{1\over 2}\,\int\,{i d^4k_1\over (2\pi)^4}\,
<q_a q_b|{\cal M}_0|k_1 k_2>\,D_0(k_1) D_0(k_2)\,
<k_1 k_2|{\cal T}|q_c q_d>\, 
\end{equation}
where $k_2=P-k_1=q_a+q_b-k_1$ and $D_0$ is the free space pion propagator.
The potential ${\cal M}_0$ given by eq.(\ref{M0})
 contains, together with the on-shell 
piece, an off-shell piece depending on the four inverse pion propagators. 
However, 
as explained in \cite{WROS}, the off-shell terms can always be absorbed 
in the potentiel by a proper redefinition of the coupling constants and mass.
It follows that the free scattering matrix can be algebraically obtained as~:
\begin{equation}
{\cal T}(s)={1\over f^2_\pi}(m^2_\pi - 2s)\,\left(
1- {1\over  2 f^2_\pi}(m^2_\pi - 2s)\,
\int{d^3 k\over (2\pi)^3}\,{v^2(k)\over \omega_k} \,
{1\over s-4\omega^2_k+i\eta}\right)^{-1} \label{SCATT0}
\end{equation}
where $v(k)$ is a form factor which can 
be fitted on low energy phase shifts and scattering length.
Evidently, an explicit sigma meson and/or a  coupling 
to the $K\bar K$ channel should be introduced to describe 
the higher energy part \cite{AOUSS, WROS}
but this is not the purpose of this paper. The question of interest here
is to study the  threshold region of this pion-pion amplitude in the nuclear 
medium. According to the approach of \cite{AOUSS}, this in-medium scattering 
matrix is obtained by simply replacing in (\ref{SCATT0}) the free pion 
propagators by the dressed one containing the p-wave pion polarizability. 
However, as pointed out in \cite{WROS}, one has to take 
care of the off-shell piece of the effective in-medium potential (\ref{MEFF})
which  contains a piece proportionnal to the p-wave polarizability. 
As claimed in \cite{WROS}, 
this yields  a contribution to the scattering amplitude,
having in the intermediate state one (dressed) pion and one (undressed)
particle-hole or delta-hole state (see fig.2c). The effect of this $\pi-ph$
configuration (calculated in a particular representation $\alpha=0$) 
was considered in \cite{WROS} yielding an accumulation 
of strength below threshold at moderate density ($\rho/\rho_0=0.5$); 
the unusual 
feature was the disparition of this effect with increasing density. 
What is also 
surprising is that this effect comes from a contribution to the in-medium 
potential (\ref{MEFF}) which is manifestly 
representation dependent. To clarify the matter we will study the first order 
contribution to the scattering amplitude 
($V^{eff}\,.\, {2\pi \,propagator}\,.\, V^{eff}$)
to first order in the polarization propagator $\Pi_0$; the various diagrams
depicted in fig. 2 are the same considered in \cite{WROS}. As in this paper 
we consider only the pieces having an imaginary part since we are 
mainly interested in the $2\pi$ strength function.
The first diagram (fig.2a ) is the usual one and
corresponds to the dressing of the pion but with the potential corrected 
by the off-shell behaviour (eq.(\ref{M0}))~:
\begin{equation}
{\cal T}^{(a)}= 2\times {1\over 2}\, \int\,{i \,d^4k_1\over (2\pi)^4}\,   
\left[V_{on} -{\beta\over f^2_\pi} 
D^{-1}_0(k_2)\right]^2 \,{\bf k}_2^2 \tilde\Pi_0(k_2) D^2_0(k_2)\,D_0(k_1) 
\end{equation}
The second diagram (fig.2b) explicitely incorporates the $4\pi$ 
vertex correction involving the polarizability $\tilde\Pi_0$~
\begin{equation}
{\cal T}^{(b)}= 4\times {1\over 2}\, \int\,{i \,d^4k_1\over (2\pi)^4}\,   
\left[V_{on} -{\beta\over f^2_\pi} \,
D^{-1}_0(k_2)\right] \,{\beta\over f^2_\pi} 
\,{\bf k}_2^2 \tilde\Pi_0(k_2) D^2_0(k_2)\,D_0(k_1) 
\end{equation}
Finally, there is a diagram (fig.2c) 
with a $p-h$ bubble in the intermediate state. 
It gives a contribution (at zero total momentum, ${\bf P}=0$)
\begin{equation}
{\cal T}^{(c)}= 2\times{1\over 2}\,\int {i d^4 k_1\over (2 \pi)^4}\,
\left({\beta\over f_\pi^2}\right)^2\,{\bf k}_2^2  \tilde\Pi^0(k_2)\, D_0(k_1)  
\end{equation}
The various prefactors come from the counting of equivalent diagrams. 
Summing this three contributions, we see that the representation
dependent piece ({\it i.e.} involving $\beta$) disappears as it should be. 
\begin{equation}
{\cal T}^{(a)} + {\cal T}^{(b)} + {\cal T}^{(c)}=
\int {i d^4 k_1\over (2\pi)^4} 
\, V^2_{on}\, {\bf k}^2_2\, \tilde\Pi_0(k_2)\,D^2_0(k_2)\,D_0(k_1)
\end{equation}   
Hence we find the ``normal'' first order dressing of the pion propagator.
This result is at variance with \cite{WROS} where the $\pi-ph$ intermediate 
state was found to contribute to the strength function. 
This result can be generalized when the pion propagator is fully dressed 
by the p-wave pion polarizability. The total second order scattering amplitude
takes the form (for ${\bf P}=0$)~:
\begin{eqnarray}
Im {\cal T}^{(2)}&=& {1\over 2}\,Im \int {i d^4 k_1
\over (2\pi)^4} \, \bigg\{\left[V_{on} -
{\beta\over f^2_\pi}\left(D^{-1}_R(k_1)+D^{-1}_R(k_2)\right)\right]^2\,
D_R(k_1)\,D_R(k_2)  \nonumber\\
&\, &\,+\left({\beta\over f^2_\pi}\right)^2\,
\left({\bf k}_1^2\,\tilde\Pi_0(k_1)\,D_R(k_2)
\,+\, {\bf k}_2^2\,\tilde\Pi_0(k_2)\,D_R(k_1) \right)\bigg\}\nonumber\\
&=& {1\over 2} \,Im\, \int {i d^4 k_1\over (2\pi)^4} 
V^2_{on}\,D_R(k_1)\,D_R(k_2)
\end{eqnarray} 
Although some more work remains to be done to write a fully unitarized 
in-medium scattering equation, we come to the conclusion that intermediate 
states with one quasi-pion and undressed $p-h$ or $\Delta-h$ states  do not
contribute to the
stength function. This is not so surprising since only fully dressed states
({\it i e.} exact eigenstates) can appear in the intermediate state of the
$\pi \pi$ scattering series.

\section{Conclusion}
Based on a chiral lagrangian we have investigated various effects
concerning pion interaction in the nuclear medium. We have in particular 
studied the interplay  between p-wave interaction responsible for 
the existence of collective pionic modes and s-wave $\pi N$ interaction.
We have insisted on the key role played by the in-medium scalar pion density
which governs a large amount of chiral symmetry restoration. Starting from 
an in-medium modified $\pi\pi$ interaction consistent with 
the p-wave dressing of the pion, we have obtained an explicit expression 
of the pion scalar density and the pion cloud contribution to the 
$\pi$-nucleus sigma term. We have shown that, contrary to the axial 
observables $f_\pi$ and $g_A$, the $\pi NN$ coupling constant remains 
unmodified to one-pion loop order. As a consequence the Goldberger-Treiman 
relation is preserved at finite density, at least in the limit 
where the effect of short-range correlations is ignored. We have 
finally demonstrated that the in-medium modification of the $\pi\pi$ 
potential cannot give genuine contribution to the two-pion strength 
function, confirming the validity of the approach developped in \cite 
{AOUSS}. This is particularly important in view of the understanding 
of the structure observed in the two-pion invariant mass spectrum in 
$A(\pi, 2\pi)$ reactions \cite{CHAOS}.
\vskip 1 true cm
{\it Acknowledgements~: We thank J. Delorme and M. Ericson for enlightening
discussions and critical reading of the manuscript.}

\vfill\eject
\begin{center}
{\huge {FIGURE CAPTIONS}}
\end{center}

Figure 1~: Vertex correction to the $\pi \pi$ interaction in the medium through
$p-h$ and $\Delta-h$ polarization bubbles which may contain the screening effect
due to short range correlation ($g'$ parameters).

Figure 2~: Medium effects appearing in $\pi \pi$ rescattering. 2a~: dressing of
the intermediate pion by the p-wave polarizability. 2b~: vertex correction to
the $\pi \pi$ potential. 2c~: contribution of the $p-h$ and $\Delta-h$
excitations in the intermediate state.

\vfill\eject

\begin{figure}
\centerline{\epsfxsize=10.5cm\epsfbox{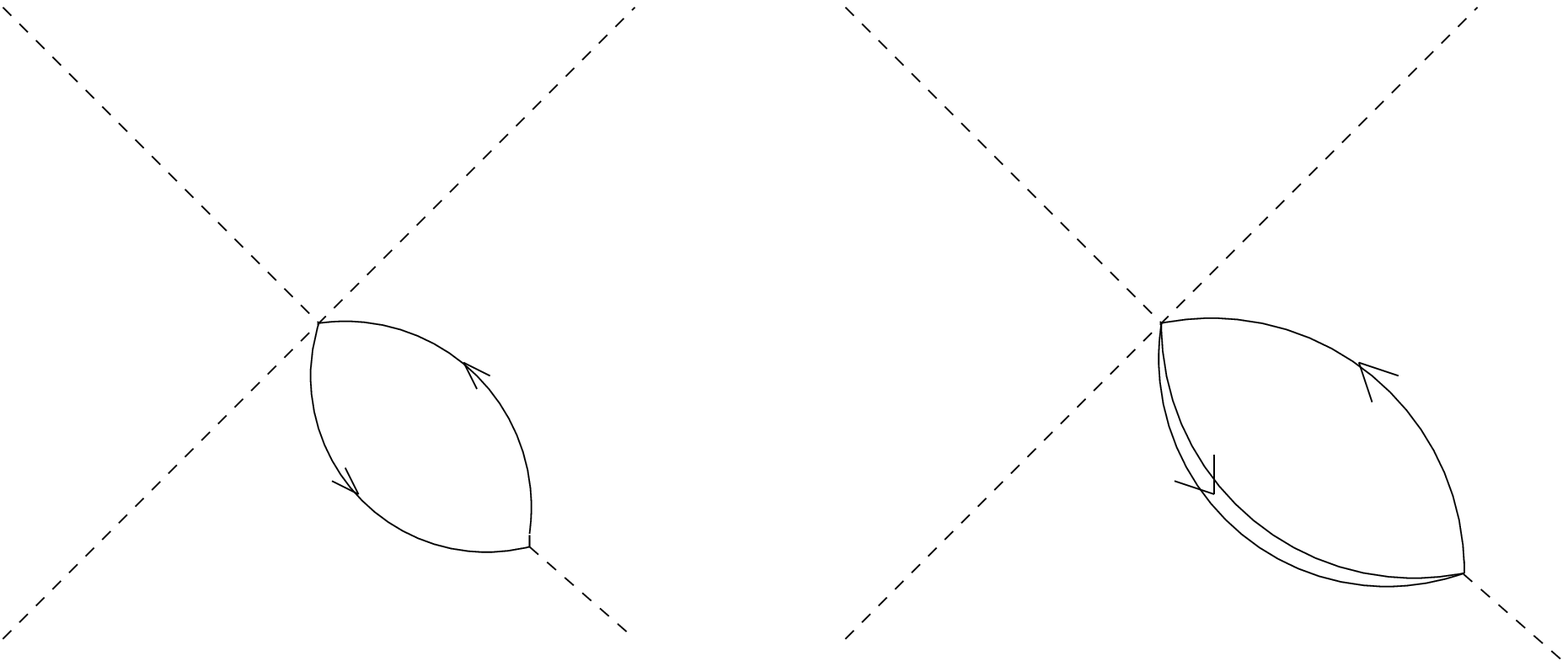}}
\end{figure}
\begin{figure}
\centerline{\epsfxsize=6.5cm\epsfbox{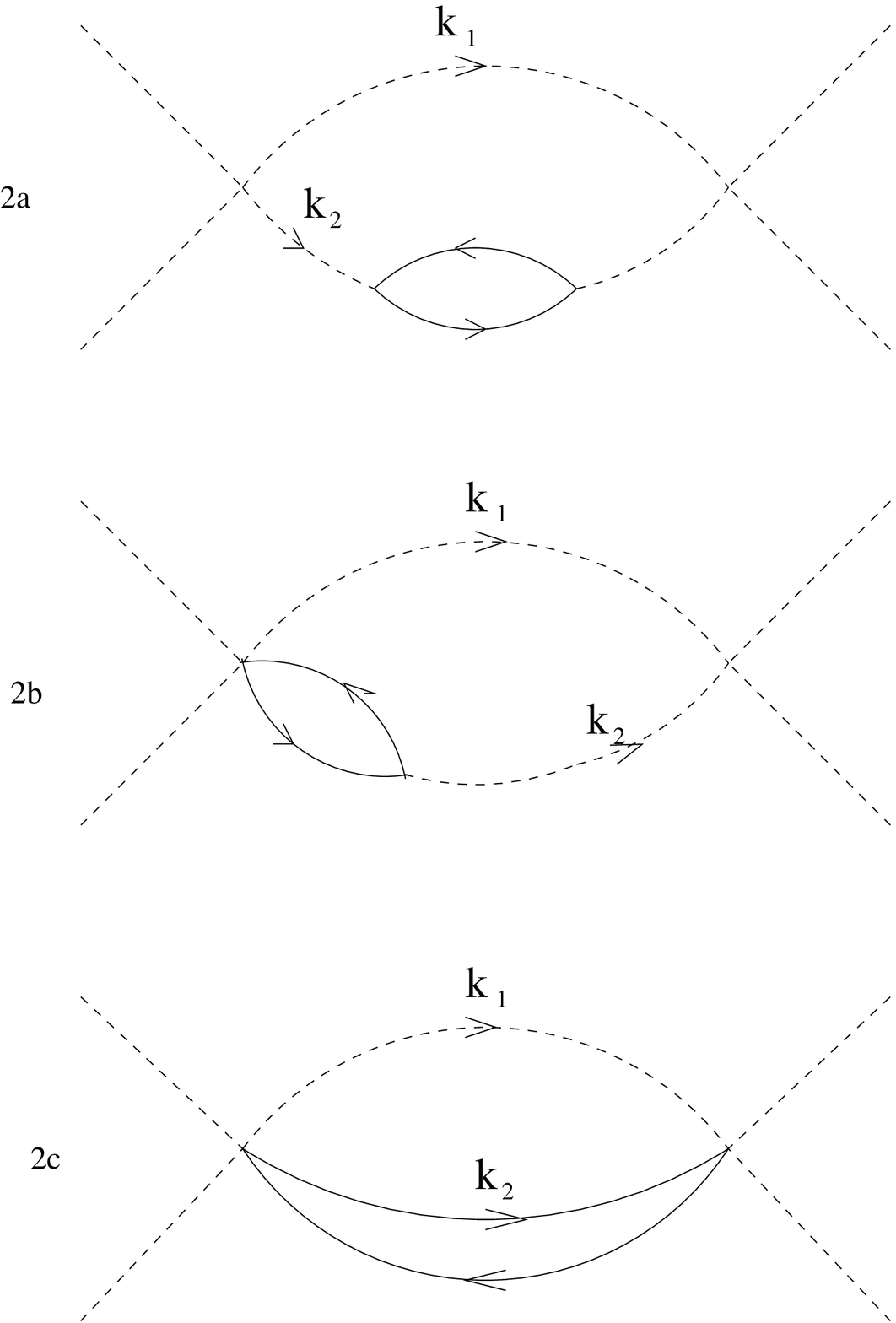}}
\end{figure}

\end{document}